\documentclass[aps, prc, floatfix, nofootinbib, superscriptaddress, twocolumn]{revtex4-1}

\usepackage{latexsym}
\usepackage{amsmath}
\usepackage{amssymb}
\usepackage{amsfonts}
\usepackage{multirow}

\usepackage[mathscr,scaled=1.15]{urwchancal}
\DeclareFontFamily{OT1}{pzc}{}
\DeclareFontShape{OT1}{pzc}{m}{it}%
{<-> s * [1.15] pzcmi7t}{}
\DeclareMathAlphabet{\mathpzc}{OT1}{pzc}{m}{it}

\usepackage{color}

\usepackage{supertabular}
\usepackage{placeins}
\usepackage{epsfig}
\usepackage{graphicx}

\definecolor{purple}{rgb}{0.5,0,0.5}
\definecolor{blue}{rgb}{0.0,0,0.9}
\definecolor{prdblue}{rgb}{0.133,0.118,0.498}
\usepackage[colorlinks=true, pdfstartview=FitV, linkcolor=prdblue, citecolor= prdblue, urlcolor=prdblue]{hyperref}

\begin{document}

\title{searches for $ud\bar{s}\bar{b}$ in the chiral quark model}
\author{Xiaoyun Chen}
\email{xychen@jit.edu.cn} \affiliation{Department of Basic
Courses, Jinling Institute of Technology, Nanjing 211169, P. R.
China}

\author{Jialun Ping}
\email[]{jlping@njnu.edu.cn} \affiliation{Department of Physics,
Nanjing Normal University, Nanjing 210023, P. R. China}

\begin{abstract}
Inspired by the report of D0 Collaboration on the state $X(5568)$
with four different flavors, a similar state, $ud\bar{s}\bar{b}$
is investigated in the present work. The advantage of looking for
this state over the state $X(5568)$ with quark contents,
$bu\bar{d}\bar{s}$ or $bd\bar{u}\bar{s}$, is that the $BK$
threshold is 270 MeV higher than that of $B_s \pi$, and it allows
a large mass region for $ud\bar{s}\bar{b}$ to be stable. The
chiral quark model and Gaussian expansion method are employed to
do the calculations of four-quark states $ud\bar{s}\bar{b}$ with
quantum numbers $IJ^{P}$($I=0,1;~ J=0,1,2;~P=+$). Two structures,
diquark-antidiquark and meson-meson, with all possible color
configurations are considered. The results indicate that energies
of the tetraquark with diquark-antiquark configuration are all
higher than the threshold of $BK$, but for the state of
$IJ^P=00^+$ in the meson-meson structure, the energies are below
the corresponding thresholds, where the color channel coupling
plays an important role. The distances between two objects
(quark/antiquark) show that the state is a molecular one.
\end{abstract}

%\pacs{14.40.Be,12.39.Pn,24.10.Eq}

\maketitle

%%%%%%%%%%%%%%%%%%%%%%%%%%%%%%%%%%%%%%%%%%%%%%%%%%%%%%%%%%%%%%%%%%%%%%%%%%%%%%%%%%%%%%%%%%%%%%%%%%%%%%%%%%%%%%%%%%%%%%%

\section{Introduction} \label{introduction}
Since the first exotic resonance $X(3872)$ was observed by the
Bell collaboration in 2003~\cite{Choi:2003}, many other exotic
sates, so called ``XYZ" states, have emerged from the reports of
Belle, BaBar, BESIII, LHCb, CDF, D0 and other collaborations. In
the traditional quark models, the meson is consist of quark and
antiquark and the baryon is made up of three quarks. How to
explain these exotic states is a big challenge for quark models.
The study of the exotic states is helpful for improving the quark
model and deepening our understanding of the nonperturbative
quantum chromodynamics (QCD).

Although, these ``XYZ" states are difficult to be explained as the
ordinary hadrons, their quantum numbers can be constructed by
quark-antiquark combinations. To find an unambiguous tetraquark
state, the particle with four different flavors are expected. Not
so long ago, the D0 Collaboration announced a new resonance named
$X(5568)$ with the mass $M=5567.8\pm2.9^{+0.9}_{-1.9}$ MeV and
narrow width $\Gamma=21.9\pm6.4^{+5.0}_{-2.5}$ MeV,
respectively~\cite{D0Co:2016}. Recently, the D0 Collaboration
reported a further evidence about this state in the weak decay of
$B$ with a significance of $6.7\sigma$ \cite{D0Co:2017}, which is
consistent with their previous measurement~\cite{D0Co:2016}.
Subsequently, searches for $X(5568)$ in decays to
$B_s^{0}\pi^{\pm}$, $B_s^{0}\rightarrow J/\psi \phi$ are performed
by the LHCb~\cite{LHCb:2016}, CMS~\cite{CMS:2017} and
ATLAS~\cite{ATLAS:2018} Collaborations in $pp$ collisions and by
the CDF Collaboration~\cite{CDF:2017} at the Tevatron, and all of
experiments revealed no signal. Clearly, more measurements are
needed.

On the theoretical side, the early work based on the QCD sum rule
supported the existence of the state
$X(5568)$~\cite{sumrule1,sumrule2,sumrule3,sumrule4,sumrule5}.
Some quark model calculations also claimed the possible
explanation of the results of the D0
Collaboration~\cite{qm1,qm2,qm3}. However, the detailed
examination of the various interpretations of the state $X(5568)$
shown that the threshold, cusp, molecular and tetraquark models
are all unfavored~\cite{Burns}. Based on the general properties of
QCD, F. K. Guo {\em et al.} argued that the QCD does not support
the existence of the state $X(5568)$~\cite{FKGuo}. Our previous
quark model calculation of tetraquark and molecule also obtained
negative results~\cite{EPJC}.

In this work, we intend to study a particle with also four
different flavors, $ud\bar{s}\bar{b}$, which is different from the
$X(5568)$ consist of $us\bar{d}\bar{b}$. For simplicity, we denote
the particle $ud\bar{s}\bar{b}$ as $T_{bs}$. The reasons for
searching this particle are as follows. Firstly, the breaking up
of $T_{bs}$ is $BK$, which the threshold is higher than the $B_s
\pi$ threshold of $X(5568)$, and it leads to a large mass region
for $T_{bs}$ to be stable. Secondly, for diquark-antidiquark
configuration, the $ud$ quark pair is more stable than the $us$
owing to the lower mass of $d$ quark than $s$ quark. In others
words, if $X(5568)$ does exist, the $T_{bs}$ must be a stable
state. If $X(5568)$ is proved to be nonexistent, there's still
probability for $T_{bs}$ to be stable. F. S. Yu also suggested
that if $ud\bar{s}\bar{b}$ exists here, the most favorable decay
mode to observe it will be $J/\psi K^-K^-\pi^+$
experimentally~\cite{yufusheng:2017}.

In order to search for the particle of $T_{bs}$ theoretically, we
calculate the masses of the states with quantum numbers $IJ^P$
($I=0,1;J=0,1,2;P=+$) including two different structures,
diquark-antidiquark and meson-meson in the chiral quark model, and
all possible color configurations are investigated by using the
Gaussian expansion method (GEM)~\cite{Hiyama:2003cu}. In the
calculation, two ways of using $\sigma$ meson exchange are
adopted. One is that the $\sigma$ meson exchange only occurs
between $u$ quark and/or $d$ quark. Another is the effective
$\sigma$ is exchanged between $u$, $d$ and $s$ quarks. If a bound
state is obtained, the average distances between quarks or
antiquarks are calculated, which can be used to clarify the
structures of the states, a compact tetraquark or a molecular
state.

The chiral quark model, the wave functions of $T_{bs}$ and the method for solving the four quark states are
detailed in Sec.\,\ref{GEM and chiral quark model}; and Sec.\,\ref{Numerical Results} is devoted to a
discussion of the results. Sec.\,\ref{epilogue} is a summary.

%%%%%%%%%%%%%%%%%%%%%%%%%%%%%%%%%%%%%%%%%%%%%%%%%%%%%%%%%%%%%%%%%%%%%%%%%%%%%%%%%%%%%%%%%%%%%%%%%%%%%%%%%%%%%%%%%%%%%%%

\section{Chiral quark model and wave functions of $T_{bs}$}
\label{GEM and chiral quark model}
\subsection{The chiral quark model}
The chiral quark model has acquired great achievements both in
describing the hadron spectra and hadron-hadron interactions. The
details of the model can be found in Ref.~\cite{Valcarce:2005em}.
Here only the Hamiltonian of the chiral quark model is given as
follows,
\begin{eqnarray}
 H & = & \sum_{i=1}^4 m_i  +\frac{p_{12}^2}{2\mu_{12}}+\frac{p_{34}^2}{2\mu_{34}}
  +\frac{p_{1234}^2}{2\mu_{1234}} +\sum_{i<j=1}^4 V_{ij},   \\
 V_{ij} & = & V_{ij}^{C}+V_{ij}^{G}+\sum_{\chi=\pi,K,\eta} V_{ij}^{\chi}
   +V_{ij}^{\sigma}.
\end{eqnarray}
The potential energy is constituted of pieces describing quark confinement (C); one-gluon-exchange (G);
one Goldstone boson exchange ($\chi=\pi$, $K$, $\eta$) and $\sigma$ exchange. Their form for the low-lying
four-quark states is \cite{Valcarce:2005em},
{\allowdisplaybreaks
\begin{subequations}
\begin{align}
V_{ij}^{C}&= ( -a_c r_{ij}^2-\Delta ) \boldsymbol{\lambda}_i^c
\cdot \boldsymbol{\lambda}_j^c ,  \\
 V_{ij}^{G}&= \frac{\alpha_s}{4} \boldsymbol{\lambda}_i^c \cdot \boldsymbol{\lambda}_{j}^c
\left[\frac{1}{r_{ij}}-\frac{2\pi}{3m_im_j}\boldsymbol{\sigma}_i\cdot
\boldsymbol{\sigma}_j
  \delta(\boldsymbol{r}_{ij})\right],  \\
\delta{(\boldsymbol{r}_{ij})} & =  \frac{e^{-r_{ij}/r_0(\mu_{ij})}}{4\pi r_{ij}r_0^2(\mu_{ij})}, \\
V_{ij}^{\pi}&= \frac{g_{ch}^2}{4\pi}\frac{m_{\pi}^2}{12m_im_j}
  \frac{\Lambda_{\pi}^2}{\Lambda_{\pi}^2-m_{\pi}^2}m_\pi v_{ij}^{\pi}
  \sum_{a=1}^3 \lambda_i^a \lambda_j^a,  \\
V_{ij}^{K}&= \frac{g_{ch}^2}{4\pi}\frac{m_{K}^2}{12m_im_j}
  \frac{\Lambda_K^2}{\Lambda_K^2-m_{K}^2}m_K v_{ij}^{K}
  \sum_{a=4}^7 \lambda_i^a \lambda_j^a,   \\
\nonumber V_{ij}^{\eta} & =
\frac{g_{ch}^2}{4\pi}\frac{m_{\eta}^2}{12m_im_j}
\frac{\Lambda_{\eta}^2}{\Lambda_{\eta}^2-m_{\eta}^2}m_{\eta}
v_{ij}^{\eta}  \\
 & \quad \times \left[\lambda_i^8 \lambda_j^8 \cos\theta_P
 - \lambda_i^0 \lambda_j^0 \sin \theta_P \right],   \\
 v_{ij}^{\chi} & =  \left[ Y(m_\chi r_{ij})-
\frac{\Lambda_{\chi}^3}{m_{\chi}^3}Y(\Lambda_{\chi} r_{ij})
\right]
\boldsymbol{\sigma}_i \cdot\boldsymbol{\sigma}_j,\\
V_{ij}^{\sigma}&= -\frac{g_{ch}^2}{4\pi}
\frac{\Lambda_{\sigma}^2}{\Lambda_{\sigma}^2-m_{\sigma}^2}m_\sigma \nonumber \\
& \quad \times \left[
 Y(m_\sigma r_{ij})-\frac{\Lambda_{\sigma}}{m_\sigma}Y(\Lambda_{\sigma} r_{ij})\right]  ,
\end{align}
\end{subequations}}
\hspace*{-0.5\parindent}%
where $Y(x)  =   e^{-x}/x$; $\{m_i\}$ is the constituent mass of $i$-th quarks/antiquarks,
and $\mu_{ij}$ is the reduced mass of the two interacting particles and
\begin{equation}
\mu_{1234}=\frac{(m_1+m_2)(m_3+m_4)}{m_1+m_2+m_3+m_4};
\end{equation}
$\mathbf{p}_{ij}=(\mathbf{p}_i-\mathbf{p}_j)/2$,
$\mathbf{p}_{1234}= (\mathbf{p}_{12}-\mathbf{p}_{34})/2$;
$r_0(\mu_{ij}) =s_0/\mu_{ij}$; $\boldsymbol{\sigma}$ are the $SU(2)$ Pauli matrices; $\boldsymbol{\lambda}$,
$\boldsymbol{\lambda}^c$ are $SU(3)$ flavor, color Gell-Mann matrices, respectively;
$g^2_{ch}/4\pi$ is the chiral coupling constant, determined from the $\pi$-nucleon coupling;
and $\alpha_s$ is an effective scale-dependent running coupling
\cite{Valcarce:2005em},
\begin{equation}
\alpha_s(\mu_{ij})=\frac{\alpha_0}{\ln\left[(\mu_{ij}^2+\mu_0^2)/\Lambda_0^2\right]}.
\end{equation}
All the parameters are determined by fitting the meson spectrum, from light to heavy; and the resulting values
are listed in Table~\ref{modelparameters}.

\begin{table}[!t]
\begin{center}
\caption{ \label{modelparameters} Model parameters, determined by
fitting the meson spectrum. }
\begin{tabular}{llr}
\hline\noalign{\smallskip}
Quark masses   &$m_u=m_d$    &313  \\
   (MeV)       &$m_s$         &536  \\
               &$m_c$         &1728 \\
               &$m_b$         &5112 \\
\hline
Goldstone bosons   &$m_{\pi}$     &0.70  \\
   (fm$^{-1} \sim 200\,$MeV )     &$m_{\sigma}$     &3.42  \\
                   &$m_{\eta}$     &2.77  \\
                   &$m_{K}$     &2.51  \\
                   &$\Lambda_{\pi}=\Lambda_{\sigma}$     &4.2  \\
                   &$\Lambda_{\eta}=\Lambda_{K}$     &5.2  \\
                   \cline{2-3}
                   &$g_{ch}^2/(4\pi)$                &0.54  \\
                   &$\theta_p(^\circ)$                &-15 \\
\hline
Confinement        &$a_c$ (MeV fm$^{-2}$)         &101 \\
                   &$\Delta$ (MeV)     &-78.3 \\
\hline
OGE                 & $\alpha_0$        &3.67 \\
                   &$\Lambda_0({\rm fm}^{-1})$ &0.033 \\
                  &$\mu_0$(MeV)    &36.98 \\
                   &$s_0$(MeV)    &28.17 \\
\hline
\end{tabular}
\end{center}
\end{table}

\subsection{The wave functions of $T_{bs}$}
We will introduce the wave functions for the two structures,
diquark-antidiquark and meson-meson, respectively. For each degree
of freedom, first we construct the wave functions for two-body
clusters, then coupling the wave functions of two clusters to the
wave functions of the tetraquark states.

(1) Diqaurk-antidiquark structure.

For spin, the wave functions for two-body clusters are,
\begin{align}
&\chi_{11}=\alpha\alpha,~~
\chi_{10}=\frac{1}{\sqrt{2}}(\alpha\beta+\beta\alpha),~~
\chi_{1-1}=\beta\beta,\nonumber \\
&\chi_{00}=\frac{1}{\sqrt{2}}(\alpha\beta-\beta\alpha),
\end{align}
then the wave functions for four-quark states are obtained,
 {\allowdisplaybreaks
\begin{subequations}\label{spinwavefunctions}
\begin{align}
\chi_{0}^{\sigma
1}&=\chi_{00}\chi_{00},\\
\chi_{0}^{\sigma
2}&=\sqrt{\frac{1}{3}}(\chi_{11}\chi_{1-1}-\chi_{10}\chi_{10}+\chi_{1-1}\chi_{11}),\\
\chi_{1}^{\sigma
3}&=\chi_{00}\chi_{11},\\
 \chi_{1}^{\sigma
4}&=\chi_{11}\chi_{00},\\
\chi_{1}^{\sigma
5}&=\frac{1}{\sqrt{2}}(\chi_{11}\chi_{10}-\chi_{10}\chi_{11}),\\
\chi_{2}^{\sigma
6}&=\chi_{11}\chi_{11},
\end{align}
\end{subequations}}
where the subscript of $\chi$ represents the total spin of $T_{bs}$, it takes the values $S=0, 1, 2$.

For flavor, the wave functions for four-quark states are
 {\allowdisplaybreaks
\begin{subequations}
\begin{align}
\chi_{d0}^{f1}&=\frac{1}{\sqrt{2}}(ud-du)\bar{s}\bar{b},\\
\chi_{d1}^{f2}&=\frac{1}{\sqrt{2}}(ud+du)\bar{s}\bar{b}.
\end{align}
\end{subequations}}
Analogously, the subscript of $\chi_d$ represents the isospin of $T_{bs}$, $I=0, 1$.

For color, there are two color configurations for the quarks pair, $[11]$ and
$[2]$, respectively.
\begin{subequations}
\begin{align}
\chi^{[11]}_1=\frac{1}{\sqrt{2}}(rg-gr),\\
\chi^{[11]}_2=\frac{1}{\sqrt{2}}(rb-br),\\
\chi^{[11]}_3=\frac{1}{\sqrt{2}}(gb-bg),
\end{align}
\end{subequations}
\begin{subequations}
\begin{align}
\chi^{[2]}_1&=rr,~~~~~ \chi^{[2]}_2=gg,
~~~~~\chi^{[2]}_3=bb,\\
\chi^{[2]}_4&=\frac{1}{\sqrt{2}}(rg+gr),\\
\chi^{[2]}_5&=\frac{1}{\sqrt{2}}(rb+br),\\
\chi^{[2]}_6&=\frac{1}{\sqrt{2}}(gb+bg).
\end{align}
\end{subequations}
For the antiquarks pair, the two color configurations are $[211]$ and $[22]$, respectively,
\begin{subequations}
\begin{align}
\chi^{[211]}_1&=\frac{1}{\sqrt{2}}(\bar{g}\bar{b}-\bar{b}\bar{g}),\nonumber \\
\chi^{[211]}_2&=\frac{1}{\sqrt{2}}(\bar{r}\bar{b}-\bar{b}\bar{r}),\nonumber\\
\chi^{[211]}_3&=\frac{1}{\sqrt{2}}(\bar{r}\bar{g}-\bar{g}\bar{r}).
\end{align}
\end{subequations}
\begin{subequations}
\begin{align}
\chi^{[22]}_1&=\bar{r}\bar{r},~~~~~\chi^{[22]}_2=\bar{g}\bar{g},~~~~~\chi^{[22]}_3=\bar{b}\bar{b},\nonumber \\
\chi^{[22]}_4&=-\frac{1}{\sqrt{2}}(\bar{g}\bar{b}+\bar{b}\bar{g}),\nonumber \\
\chi^{[22]}_5&=\frac{1}{\sqrt{2}}(\bar{r}\bar{b}+\bar{b}\bar{r}),\nonumber\\
\chi^{[22]}_6&=-\frac{1}{\sqrt{2}}(\bar{r}\bar{g}+\bar{g}\bar{r}).
\end{align}
\end{subequations}
The color wave functions of $T_{bs}$ in the diquark-antidiquark
structure should be color singlet $[222]$ and it can be obtained
by using $SU(3)$ Clebsh-Gordan coefficients,
\begin{subequations}
\begin{align}
\chi^{c}_{d1} & =
\frac{\sqrt{3}}{6}(rg\bar{r}\bar{g}-rg\bar{g}\bar{r}+gr\bar{g}\bar{r}-gr\bar{r}\bar{g} \nonumber \\
&~~~+rb\bar{r}\bar{b}-rb\bar{b}\bar{r}+br\bar{b}\bar{r}-br\bar{r}\bar{b} \nonumber \\
&~~~+gb\bar{g}\bar{b}-gb\bar{b}\bar{g}+bg\bar{b}\bar{g}-bg\bar{g}\bar{b}).  \\
\chi^{c}_{d2}&=\frac{\sqrt{6}}{12}(2rr\bar{r}\bar{r}+2gg\bar{g}\bar{g}+2bb\bar{b}\bar{b}
    +rg\bar{r}\bar{g}+rg\bar{g}\bar{r} \nonumber \\
&~~~+gr\bar{g}\bar{r}+gr\bar{r}\bar{g}+rb\bar{r}\bar{b}+rb\bar{b}\bar{r}+br\bar{b}\bar{r} \nonumber \\
&~~~+br\bar{r}\bar{b}+gb\bar{g}\bar{b}+gb\bar{b}\bar{g}+bg\bar{b}\bar{g}+bg\bar{g}\bar{b}).
\end{align}
\end{subequations}
Where, $\chi_{d1}^{c}$ and $\chi_{d2}^{c}$ represents the color
antitriplet-triplet ($\bar{3}\times3$) and sextet-antisextet
($6\times\bar{6}$), respectively.

(2)Meson-meson structure.

For spin, the wave functions are the same as those of the
diquark-antidiquark structure, Eq.~(\ref{spinwavefunctions}).

The flavor wave functions of $T_{bs}$ take as follows,
\begin{subequations}
\begin{align}
\chi_{m0}^{f1}&=\frac{1}{\sqrt{2}}(\bar{s}u\bar{b}d-\bar{s}d\bar{b}u),\\
\chi_{m1}^{f2}&=\frac{1}{\sqrt{2}}(\bar{s}u\bar{b}d+\bar{s}d\bar{b}u),
\end{align}
\end{subequations}
the subscript of $\chi_m$ represents the isospin of $T_{bs}$, $I=0, 1$.

For color wave functions, the possible color configurations of a cluster consist of a quark and
an antiquark are $[111]$, and $[21]$,
\begin{align}
\chi^{[111]}_1=\frac{1}{\sqrt{3}}(\bar{r}r+\bar{g}g+\bar{b}b).
\end{align}
\begin{align}
&\chi^{[21]}_1=\bar{b}r,~~~~~\chi^{[21]}_2=\bar{b}g,~~~~~\chi^{[21]}_3=-\bar{g}r, \nonumber \\
&\chi^{[21]}_4=\frac{1}{\sqrt{2}}(\bar{r}r-\bar{g}g),~~~~\chi^{[21]}_5=\frac{1}{\sqrt{6}}(2\bar{b}b-\bar{r}r-\bar{g}g),\nonumber \\
&\chi^{[21]}_6=\bar{r}g,~~~~~\chi^{[21]}_7=-\bar{g}b,~~~~~\chi^{[21]}_8=\bar{r}b.
\end{align}
At last, the color singlet wave functions of $T_{bs}$ in the meson-meson structure are,
\begin{subequations}
\begin{align}
\chi_{m1}^{c}&=\frac{1}{3}(\bar{r}r+\bar{g}g+\bar{b}b)(\bar{r}r+\bar{g}g+\bar{b}b),\\
\chi_{m2}^{c}&=\frac{\sqrt{2}}{12}(3\bar{b}r\bar{r}b+3\bar{g}r\bar{r}g+3\bar{b}g\bar{g}b+3\bar{g}b\bar{b}g+3\bar{r}g\bar{g}r \nonumber \\
&~~~+3\bar{r}b\bar{b}r+2\bar{r}r\bar{r}r+2\bar{g}g\bar{g}g+2\bar{b}b\bar{b}b-\bar{r}r\bar{g}g \nonumber\\
&~~~-\bar{g}g\bar{r}r-\bar{b}b\bar{g}g-\bar{b}b\bar{r}r-\bar{g}g\bar{b}b-\bar{r}r\bar{b}b).
\end{align}
\end{subequations}
Where, $\chi_{m1}^{c}$ and $\chi_{m2}^{c}$ represents the color
singlet-singlet($1\times1$) and color octet-octet($8\times8$),
respectively.

 As for the spatial
wave functions, the total orbital wave functions can be
constructed by coupling the orbital wave function for each
relative motion of the system,
\begin{align}\label{spatialwavefunctions}
\Psi_{L}^{M_{L}}=\left[[\Psi_{l_1}({\bf r}_{12})\Psi_{l_2}({\bf r}_{34})]_{l_{12}}\Psi_{L_r}({\bf r}_{1234})
\right]_{L}^{M_{L}},
\end{align}
where $L$ is the total orbital angular momentum of $T_{bs}$ and
$\Psi_{L_r}(\mathbf{r}_{1234})$ is the wave function of the
relative motion between two sub-clusters with orbital angular
momentum $L_r$, and the Jacobi coordinates are defined as,
\begin{align}\label{jacobi}
{\bf r}_{12}&={\bf r}_1-{\bf r}_2, \nonumber \\
{\bf r}_{34}&={\bf r}_3-{\bf r}_4, \nonumber\\
{\bf r}_{1234}&=\frac{m_1{\bf r}_1+m_2{\bf
r}_2}{m_1+m_2}-\frac{m_3{\bf r}_3+m_4{\bf r}_4}{m_3+m_4},
\end{align}
for diquark-antidiquark structure, the quarks are numbered as $1,
2$, and the antiquarks are numbered as $3, 4$; for meson-meson
structure, one cluster with antiquark and quark is marked as $1,
2$, the other cluster with antiquark and quark is marked as $3,
4$. In GEM, the spatial wave function is expanded by
Gaussians~\cite{Hiyama:2003cu}:
\begin{subequations}
\label{radialpart}
\begin{align}
\Psi_{l}^{m}(\mathbf{r}) & = \sum_{n=1}^{n_{\rm max}} c_{n}\psi^G_{nlm}(\mathbf{r}),\\
\psi^G_{nlm}(\mathbf{r}) & = N_{nl}r^{l}
e^{-\nu_{n}r^2}Y_{lm}(\hat{\mathbf{r}}),
\end{align}
\end{subequations}
where $N_{nl}$ are normalization constants,
\begin{align}
N_{nl}=\left[\frac{2^{l+2}(2\nu_{n})^{l+\frac{3}{2}}}{\sqrt{\pi}(2l+1)}
\right]^\frac{1}{2}.
\end{align}
$c_n$ are the variational parameters, which are determined
dynamically. The Gaussian size parameters are chosen according to
the following geometric progression
\begin{equation}\label{gaussiansize}
\nu_{n}=\frac{1}{r^2_n}, \quad r_n=r_1a^{n-1}, \quad
a=\left(\frac{r_{n_{\rm max}}}{r_1}\right)^{\frac{1}{n_{\rm
max}-1}}.
\end{equation}
This procedure enables optimization of the ranges using just a
small number of Gaussians. Finally, the complete channel wave
function for the four-quark system for diquark-antidiquark
structure is written as
\begin{align}\label{diquarkpsi}
&\Psi_{IJ,i,j,k}^{M_IM_J}=[\Psi_{L}\chi_S^{\sigma i}]_{J}^{M_J}\chi_{dI}^{fj}\chi^{c}_{dk},\nonumber \\
&(i=1\sim6, j=1,2, k=1,2; S=0,1,2; I=0,1).
\end{align}
For meson-meson structure, the complete wave function is written as
\begin{align}
&\Psi_{IJ,i,j,k}^{M_IM_J}= {\cal A}[\Psi_{L}\chi_S^{\sigma
i}]_{J}^{M_J}\chi_{mI}^{fj}\chi^{c}_{mk},\nonumber \label{mesonpsi}\\
&(i=1\sim6, j=1,2, k=1,2; S=0,1,2; I=0,1).
\end{align}
Here, ${\cal A}$ is the antisymmetrization operator: if all quarks (antiquarks) are taken as
identical particles, then
\begin{equation}
{\cal A}=\frac{1}{2}(1-P_{13}-P_{24}+P_{13}P_{24}).
\end{equation}
In the present work, for $T_{bs}$ system, only two quarks are the
identical particles, so the antisymmetrization operator used is
\begin{equation}
{\cal A}=\frac{1}{\sqrt{2}}(1-P_{13}).
\end{equation}

The eigenenergy of the $T_{bs}$  system is obtained by solving a Schr\"{o}dinger equation:
\begin{equation}
    H \, \Psi^{\,M_IM_J}_{IJ}=E^{IJ} \Psi^{\,M_IM_J}_{IJ},
\end{equation}
where $\Psi^{\,M_IM_J}_{IJ}$ is the wave function of the $T_{bs}$, which is the linear combinations
of the above channel wave functions, Eq.~(\ref{diquarkpsi}) in the diquark-antidiquark
structure or Eq.~(\ref{mesonpsi}) in the meson-meson structure, respectively.

The calculation of Hamiltonian matrix elements is complicated if any one of the relative orbital
angular momenta is nonzero. In this case, it is useful to employ the method of infinitesimally shifted
Gaussians \cite{Hiyama:2003cu}, wherewith the spherical harmonics are absorbed into the Gaussians:
\begin{eqnarray}
\label{wavefunctionG}
 \psi^G_{nlm}(\mathbf{r})& =& N_{nl}r^{l}
 e^{-\nu_{n}r^2}Y_{lm}(\hat{\mathbf{r}}) \nonumber \\
 & =&N_{nl}\lim_{\epsilon \rightarrow
 0}\frac{1}{\epsilon^{l}}\sum_{k}^{k_{\rm max}}C_{lm,k} \, {\rm e}^{-\nu_{n}(\mathbf{r}-\epsilon\mathbf{D_{lm,k}})^2},
\end{eqnarray}
where, plainly, the quantities $C_{lm,k}$, $D_{lm,k}$ are fixed by the particular spherical harmonic under consideration
and their values ensure the limit $\epsilon\to 0$ exists.

%%%%%%%%%%%%%%%%%%%%%%%%%%%%%%%%%%%%%%%%%%%%%%%%%%%%%%%%%%%%%%%%%%%%%%%%%%%%%%%%%%%%%%%%%%%%%%%%%%%%%%%%%%%%%%%%%%%%%%%

\section{Numerical Results and discussions}
\label{Numerical Results}

In the present work, we try to search the particle with quantum
numbers $IJ^P(I=0,1; J=0, 1, 2; P=+$) consist of four different
flavors $ud\bar{s}\bar{b}$, denoted as $T_{bs}$, in the chiral
quark model. All the orbital angular momenta are set to zero because we are
interested in the low-lying states. The chiral quark model gives a good
description of the meson spectrum which can seen in the comparison of
theoretical thresholds and experimental thresholds in Table \ref{resultsu2diquark}.
Two structures of $T_{bs}$, diquark-antidiquark and
meson-meson, are investigated. In each structure, all possible
states are considered. For diquark-antidiquark structure, two
color configurations, color
antitriplet-triplet ($\bar{3}$$\times$3) and
sexet-antisextet (6$\times$$\bar{6}$) are examined. And for
meson-meson structure, color singlet-singlet (1$\times$1) and
octet-octet (8$\times$8) are taken into account.

In $SU(2)$ flavor symmetry, the $\sigma$ meson exchange only
occurs between $u$ quark and $d$ quark. In this situation, the
results of $T_{bs}$ for diquark-antidiquark and meson-meson
structure, are given in Tables \ref{resultsu2diquark} and
\ref{resultsu2meson}, respectively.

\begin{table}[!t]
\begin{center}
\caption{ \label{resultsu2diquark} The eigenenergies of $T_{bs}$
in $SU(2)$ flavor symmetry for diquark-antidiquark
structure (unit: MeV). $E_{th1}$ represents the theoretical threshold
and $E_{th2}$ denotes the experimental threshold.}
\begin{tabular}{cccccc} \hline
$IJ^P$~~~ &channel ~~~&$E_s$~~~&$E_{cc}$~~~&$E_{th1}$~~~ &$E_{th2}$\\
 \hline
$00^+$ &$\chi_{0}^{\sigma1}\chi_{d0}^{f1}\chi_{d1}^{c}$ &6033.8 &6012.9 &5774.9 &5773.3 \\
       &$\chi_{0}^{\sigma2}\chi_{d0}^{f1}\chi^{c}_{d2}$ &6370.4 & &6233.2  &6216.9 \\
       \hline
$01+$  &$\chi_{1}^{\sigma3}\chi_{d0}^{f1}\chi^{c}_{d1}$ &6048.5 &6037.0 &5813.6 &5818.9 \\
       &$\chi_{1}^{\sigma4}\chi_{d0}^{f1}\chi^{c}_{d2}$ &6480.3 & &6194.5  &6171.3 \\
       &$\chi_{1}^{\sigma5}\chi_{d0}^{f1}\chi^{c}_{d2}$ &6426.3 & &6233.2  &6216.9 \\
       \hline
$02+$  &$\chi_{2}^{\sigma6}\chi_{d0}^{f1}\chi^{c}_{d2}$ &6522.9 &6522.9 &6233.2  &6216.9 \\
\hline
$10+$  &$\chi_{0}^{\sigma1}\chi_{d1}^{f2}\chi^{c}_{d2}$ &6514.2 &6383.5 &5774.9 &5773.3 \\
       &$\chi_{0}^{\sigma2}\chi_{d1}^{f2}\chi^{c}_{d1}$ &6417.9 & &6233.2  &6216.9 \\
       \hline
$11+$  &$\chi_{1}^{\sigma3}\chi_{d1}^{f2}\chi^{c}_{d2}$ &6510.7 &6396.8 &5813.6  &5818.9 \\
       &$\chi_{1}^{\sigma4}\chi_{d1}^{f2}\chi^{c}_{d1}$ &6448.7 & &6194.5  &6171.3 \\
       &$\chi_{1}^{\sigma5}\chi_{d1}^{f2}\chi^{c}_{d1}$ &6440.6 & &6233.2  &6216.9 \\
       \hline
$12+$  &$\chi_{2}^{\sigma6}\chi_{d1}^{f2}\chi^{c}_{d1}$ &6482.8 &6482.8 &6233.2  &6216.9 \\
  \hline
\end{tabular}
\end{center}
\end{table}

\begin{table}[!t]
\begin{center}
\caption{ \label{resultsu2meson} The eigenenergies of $T_{bs}$ in
$SU(2)$ flavor symmetry for meson-meson structure (unit:MeV). }
\begin{tabular}{ccccccc} \hline
$IJ^P$~&channel~&$E_s$&$E_{cc1}$~&$E_{cc2}$~&$E_{cc3}$~ &$E_{th1}$ \\
\hline
$00^+$ &$\chi^{\sigma 1}_{0}\chi^{f1}_{m0}\chi^{c}_{m1}$  &5775.3 &5774.4 &5774.4 &5774.3 &5774.9     \\
       &$\chi^{\sigma 1}_{0}\chi^{f1}_{m0}\chi^{c}_{m2}$  &6472.7 &       &       &       &               \\
       &$\chi^{\sigma 2}_{0}\chi^{f1}_{m0}\chi^{c}_{m1}$  &6234.7 &6222.9 &       &       &6233.2     \\
       &$\chi^{\sigma 2}_{0}\chi^{f1}_{m0}\chi^{c}_{m2}$  &6269.5 &       &       &       &               \\
       \hline
$01^+$ &$\chi^{\sigma 3}_{1}\chi^{f1}_{m0}\chi^{c}_{m1}$  &5814.2 &5813.8 &5813.8 &5813.7 &5813.6      \\
       &$\chi^{\sigma 3}_{1}\chi^{f1}_{m0}\chi^{c}_{m2}$  &6478.1 &       &       &       &               \\
       &$\chi^{\sigma 4}_{1}\chi^{f1}_{m0}\chi^{c}_{m1}$  &6195.7 &6195.3 &       &       &6194.5      \\
       &$\chi^{\sigma 4}_{1}\chi^{f1}_{m0}\chi^{c}_{m2}$  &6475.5 &       &       &       &               \\
       &$\chi^{\sigma 5}_{1}\chi^{f1}_{m0}\chi^{c}_{m1}$  &6234.8 &6233.8 &       &       &6233.2      \\
       &$\chi^{\sigma 5}_{1}\chi^{f1}_{m0}\chi^{c}_{m2}$  &6373.4 &       &       &       &               \\
       \hline
$02^+$ &$\chi^{\sigma 6}_{2}\chi^{f1}_{m0}\chi^{c}_{m1}$  &6234.0 &6234.0 &6234.0 &6234.0 &6233.2      \\
       &$\chi^{\sigma 6}_{2}\chi^{f1}_{m0}\chi^{c}_{m2}$  &6601.8 &       &       &       &               \\
       \hline
$10^+$
       &$\chi^{\sigma 1}_{0}\chi^{f2}_{m1}\chi^{c}_{m1}$  &5776.9 &5776.9 &5776.9 &5776.9 &5774.9      \\
       &$\chi^{\sigma 1}_{0}\chi^{f2}_{m1}\chi^{c}_{m2}$  &6531.6 &       &       &       &               \\
       &$\chi^{\sigma 2}_{0}\chi^{f2}_{m1}\chi^{c}_{m1}$  &6235.0 &6235.0 &       &       &6233.2      \\
       &$\chi^{\sigma 2}_{0}\chi^{f2}_{m1}\chi^{c}_{m2}$  &6481.4 &       &       &       &               \\
       \hline
$11+$  &$\chi^{\sigma 3}_{1}\chi^{f2}_{m1}\chi^{c}_{m1}$  &5815.7 &5815.7 &5815.7 &5815.7 &5813.6      \\
       &$\chi^{\sigma 3}_{1}\chi^{f2}_{m1}\chi^{c}_{m2}$  &6523.3 &       &       &       &               \\
       &$\chi^{\sigma 4}_{1}\chi^{f2}_{m1}\chi^{c}_{m1}$  &6196.6 &6196.6 &       &       &6194.5      \\
       &$\chi^{\sigma 4}_{1}\chi^{f2}_{m1}\chi^{c}_{m2}$  &6520.0 &       &       &       &               \\
       &$\chi^{\sigma 5}_{1}\chi^{f2}_{m1}\chi^{c}_{m1}$  &6235.2 &6235.2 &       &       &6233.2     \\
       &$\chi^{\sigma 5}_{1}\chi^{f2}_{m1}\chi^{c}_{m2}$  &6493.9 &       &       &       &              \\
       \hline
$12+$  &$\chi^{\sigma 6}_{2}\chi^{f2}_{m1}\chi^{c}_{m1}$  &6235.5 &6235.5 &6235.5 &6235.5 &6233.2     \\
       &$\chi^{\sigma 6}_{2}\chi^{f2}_{m1}\chi^{c}_{m2}$  &6532.7 &       &       &       &              \\
\hline
\end{tabular}
\end{center}
\end{table}

In Table \ref{resultsu2diquark}, the second column gives the
index of the antisymmetry wave functions of $T_{bs}$.
$E_s$ is the single channel eigenenergy for the
different channels; $E_{cc}$ represents the eigenenergy with the effect of
channel-coupling of different spin-color configurations.
%, antitriplet-triplet
%($\bar{3}\times3$) and sextet-antisextet($6\times\bar{6}$).
%From Table \ref{resultsu2diquark}, we can see that the theoretical
%thresholds of states in the framework of the chiral quark model
%are consistent with those experimentally.
From the table, we can see that the channels with different spin-color configurations
have similar energies and the coupling of them is
rather strong. However, the energies are all higher than the threshold
of $BK$, 5773MeV, which indicates that there are no bound states
with the diquark-antidiquark structure in our model calculation.

\begin{figure}[t]
\centerline{\includegraphics[scale=0.35]{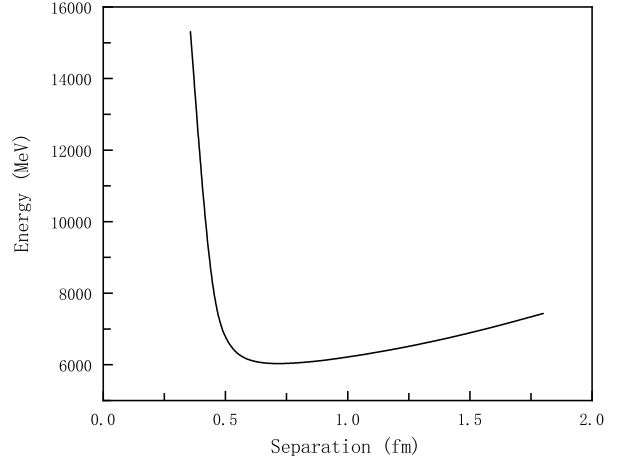}}
\caption{\label{veff} The eigenenergy of $00^+$ state as a
function of the distance between the diquark and antidiquark.}
\end{figure}
Because of the color structure, colorful sub-clusters cannot fall
apart, there may be a resonance even though the higher energy of
the state. To check the possibility, we calculate the variety of
the eigenenergy of $00^+$ state with the distance between the
diquark and antidiquark sub-clusters and the results are shown in
Fig~\ref{veff}. In this case the number of gaussians used for the
relative motion of the two sub-clusters is set to 1. we can see
that when the two sub-clusters approach closely or fall apart, the
energy is increasing, the minimum of the energy occurs around the
separation 0.6 fm. The results indicate that the two sub-clusters
cannot fall apart or get too close. So that the state turned to be
two mesons $B$ and $K$ is hindered because of the separation.
$00^+$ state may be a resonance state in our present calculation.

 With regard to meson-meson structure, the results are shown in Table
\ref{resultsu2meson}.  In our calculations, the color
singlet-singlet configurations always have the lower energies than
those of color octet-octet ones. $E_{cc1}$ is the eigenenergy from
the channel coupling of the two color configurations, which is
close to that of single channel (color singlet-singlet) result,
$E_s$. This indicates that the effect of the hidden color is very
small. $E_{cc2}$ gives the eigen-energy from the channel coupling
of all the color singlet-singlet ones, and the results show that
the coupling is also very small. $E_{cc3}$ represents the
eigen-energy from the channel coupling of all channels with the
same quantum numbers. Naturally, the coupling tends to be small.
The obtained energies $E_{cc3}$ are all higher and approach to the
theoretical thresholds in all case except the state, $00^+$. For
$IJ^P=00^+$ state, the eigen-energies from single channel
calculation are higher than their theoretical thresholds. With the
help of channel coupling to the color octet-octet configuration,
the energies of the states are lower than their corresponding
thresholds. For the first channel with spin $0\times0 \rightarrow
0$, the calculated energy is $5774.4$ MeV, which is lower than the
theoretical threshold 5774.9 MeV, and The binding energy is $-0.5$
MeV. For the second channel with spin $1\times1 \rightarrow 0$,
the obtained energy is $6222.9$ MeV, a little smaller than the
theoretical threshold 6233.2 MeV, and the binding energy is
$-10.3$ MeV. All channels coupling obtain the lowest state with
binding energy $-0.6$ MeV and push $1\times1 \rightarrow 0$ state
above its corresponding threshold. To identify which terms in the
Hamiltonian making the state be bound, the contributions from each
term of Hamiltonian for the four-quark state and the sum of two
mesons are given in Table \ref{contribution1} for both two $00^+$
states. From the table, we can see that the color confinement, one
gluon exchange and $\sigma$-meson exchange contribute the binding
of the states. For spin $1\times 1 \rightarrow 0$ state,
$\pi$-meson exchange makes a considerable binding due to the
compact structure of the state (see Table \ref{rms1}). For spin
$0\times 0 \rightarrow 0$ state, $\pi$-meson exchange makes no
contribution because of the large separation between two mesons.
F. Close $\emph{et al.}$ also found that the pion exchange between
hadrons can lead to deeply bound quasimolecular
states~\cite{Close:2009,Close:2010}.

\begin{table}[!t]
\begin{center}
\caption{ \label{contribution1} The contributions of each term of
the Hamiltonian for $00^+$ state in meson-meson structure in
$SU(2)$ flavor symmetry (unit:MeV). $\Delta_i (i=1,2)$ is the
difference between the contributions in four-quark state and the
sum of the contributions of two mesons.}
\begin{tabular}{ccccccc} \hline
\multirow{2}*{\,} &\multicolumn{3}{c}{$0\times0 \rightarrow 0$} &\multicolumn{3}{c}{$1\times1 \rightarrow 0$}\\
\hline &$T_{bs}$  & $BK$  &$\Delta_1$  &$T_{bs}$ & $B^*K^*$ &$\Delta_2$ \\
\hline rest mass &6274.0 &6274.0 &0 &6274.0 &6274.0 &0\\
\hline
kinetic &1508.4 &1493.4 &15  &966.8 &740.3 &226.5\\
\hline
$V_{ij}^G$ &-1464.1 &-1454.1 &-10 &-615.4 &-488.9 &-126.5 \\
\hline
$V_{ij}^C$ &-476.7 &-473.1 &-3.6 &-317.4 &-295.4 &-22.0\\
\hline
$V_{ij}^{\eta}$  &-65.8 &-65.2 &-0.6 &8.4 &3.3 &5.1\\
\hline
 $V_{ij}^{\pi}$  &1.4 &0.0 &1.4 &-77.0 &0.0 &-77.0\\
\hline
$V_{ij}^{K}$  &0 &0 &0 &0 &0 &0\\
\hline
$V_{ij}^{\sigma}$  &-2.8 &-0.007 &-2.7 &-16.5 &-0.002 &-16.5\\
\hline eigenenergy &5774.4 &5774.9 &-0.5 &6222.9 &6233.2 &-10.3\\
\hline
\end{tabular}
\end{center}
\end{table}

\begin{table}[!t]
\begin{center}
\caption{ \label{rms1} the RMS distances between quarks and antiquarks for the state $00^+$
in meson-meson structure in $SU(2)$ flavor symmetry (unit:fm). }
\begin{tabular}{ccccccc} \hline
 channel ~~~~&~~$u\bar{s}$  ~~~~&$d\bar{b}$  ~~~~&$ud$  ~~~~&$\bar{s}\bar{b}$  ~~~~&$u\bar{b}$  ~~~~&$\bar{s}d$ \\ \hline
 $0\times0 \rightarrow 0$ &0.5 &0.6 &6.1 &6.1 &6.1 &6.1 \\
 \hline
 $1\times1 \rightarrow 0$ &0.8 &0.6 &1.4 &1.2 &1.2 &1.3 \\
 \hline
 coupling     &0.5 &0.6 &5.6 &5.5 &5.5 &5.5\\
 \hline
\end{tabular}
\end{center}
\end{table}

\begin{table}[!t]
\begin{center}
\caption{ \label{resultsu3diquark} The eigenenergies of $T_{bs}$
in $SU(3)$ flavor symmetry for diquark-antidiquark
structure (unit: MeV). }
\begin{tabular}{ccccccc} \hline
$IJ^P$ ~~~&channel ~~~&$E_s$~~~&$E_{cc}$~~~ &$E_{th1}$~~~ &$E_{th2}$ \\
 \hline
$00^+$ &$\chi_{0}^{\sigma1}\chi_{d0}^{f1}\chi^{c}_{d1}$ &5954.0 &5931.5 &5774.9  &5773.3 \\
       &$\chi_{0}^{\sigma2}\chi_{d0}^{f1}\chi^{c}_{d2}$ &6322.9 & &6233.2  &6216.9 \\
       \hline
$01+$  &$\chi_{1}^{\sigma3}\chi_{d0}^{f1}\chi^{c}_{d1}$ &5968.5 &5956.6 &5813.6  &5818.9 \\
       &$\chi_{1}^{\sigma4}\chi_{d0}^{f1}\chi^{c}_{d2}$ &6438.7 & &6194.5  &6171.3 \\
       &$\chi_{1}^{\sigma5}\chi_{d0}^{f1}\chi^{c}_{d2}$ &6381.9 & &6233.2  &6216.9 \\
       \hline
$02+$  &$\chi_{2}^{\sigma6}\chi_{d0}^{f1}\chi^{c}_{d2}$ &6483.2 &6483.2 &6233.2  &6216.9 \\
\hline
$10+$  &$\chi_{0}^{\sigma1}\chi_{d1}^{f2}\chi^{c}_{d2}$ &6476.2 &6332.8 &5774.9 &5773.3 \\
       &$\chi_{0}^{\sigma2}\chi_{d1}^{f2}\chi^{c}_{d1}$ &6367.5 & &6233.2  &6216.9 \\
       \hline
$11+$  &$\chi_{1}^{\sigma3}\chi_{d1}^{f2}\chi^{c}_{d2}$ &6472.5 &6346.8 &5813.6  &5818.9 \\
       &$\chi_{1}^{\sigma4}\chi_{d1}^{f2}\chi^{c}_{d1}$ &6400.0 & &6194.5  &6171.3 \\
       &$\chi_{1}^{\sigma5}\chi_{d1}^{f2}\chi^{c}_{d1}$ &6391.4 & &6233.2  &6216.9 \\
       \hline
$12+$  &$\chi_{2}^{\sigma6}\chi_{d1}^{f2}\chi^{c}_{d1}$ &6435.5 &6435.5 &6233.2 &6216.9 \\
  \hline
\end{tabular}
\end{center}
\end{table}

\begin{table}[!t]
\begin{center}
\caption{ \label{resultsu3meson} The eigenenergies of $T_{bs}$ in
$SU(3)$ flavor symmetry for meson-meson structure (unit: MeV).
$E_{b}$ represents the binding energy of states.}
\begin{tabular}{cccccccc} \hline
$IJ^P$&channel&$E_s$ &$E_{cc1}$ &$E_{cc2}$ &$E_{cc3}$ &$E_{th1}$ &$E_{b}$ \\
\hline
$00^+$ &$\chi^{\sigma 1}_{0}\chi^{f1}_{m0}\chi^{c}_{m1}$  &5707.5 &5704.7 &5704.6 &5704.2 &5774.9 &-70.2    \\
       &$\chi^{\sigma 1}_{0}\chi^{f1}_{m0}\chi^{c}_{m2}$  &6420.2 &       &       &       &       &        \\
       &$\chi^{\sigma 2}_{0}\chi^{f1}_{m0}\chi^{c}_{m1}$  &6201.5 &6168.7 &       &       &6233.2 &-64.5    \\
       &$\chi^{\sigma 2}_{0}\chi^{f1}_{m0}\chi^{c}_{m2}$  &6206.8 &       &       &       &       &        \\
       \hline
$01^+$ &$\chi^{\sigma 3}_{1}\chi^{f1}_{m0}\chi^{c}_{m1}$  &5747.3 &5745.6 &5745.5 &5745.2 &5813.6 &-68     \\
       &$\chi^{\sigma 3}_{1}\chi^{f1}_{m0}\chi^{c}_{m2}$  &6425.9 &       &       &       &       &        \\
       &$\chi^{\sigma 4}_{1}\chi^{f1}_{m0}\chi^{c}_{m1}$  &6164.3 &6162.2 &       &       &6194.5 &-32.3     \\
       &$\chi^{\sigma 4}_{1}\chi^{f1}_{m0}\chi^{c}_{m2}$  &6423.5 &       &       &       &       &        \\
       &$\chi^{\sigma 5}_{1}\chi^{f1}_{m0}\chi^{c}_{m1}$  &6203.6 &6196.9 &       &       &6233.2 &-36.3     \\
       &$\chi^{\sigma 5}_{1}\chi^{f1}_{m0}\chi^{c}_{m2}$  &6316.2 &       &       &       &       &        \\
       \hline
$02^+$ &$\chi^{\sigma 6}_{2}\chi^{f1}_{m0}\chi^{c}_{m1}$  &6202.3 &6202.2 &6202.2 &6202.2 &6233.2 &-31     \\
       &$\chi^{\sigma 6}_{2}\chi^{f1}_{m0}\chi^{c}_{m2}$  &6554.7 &       &       &       &       &        \\
       \hline
$10^+$
       &$\chi^{\sigma 1}_{0}\chi^{f2}_{m1}\chi^{c}_{m1}$  &5713.6 &5713.6 &5713.6 &5713.6 &5774.9 &-61.3     \\
       &$\chi^{\sigma 1}_{0}\chi^{f2}_{m1}\chi^{c}_{m2}$  &6483.9 &       &       &       &       &        \\
       &$\chi^{\sigma 2}_{0}\chi^{f2}_{m1}\chi^{c}_{m1}$  &6204.7 &6204.7 &       &       &6233.2 &-28.5     \\
       &$\chi^{\sigma 2}_{0}\chi^{f2}_{m1}\chi^{c}_{m2}$  &6429.5 &       &       &       &       &        \\
       \hline
$11+$  &$\chi^{\sigma 3}_{1}\chi^{f2}_{m1}\chi^{c}_{m1}$  &5752.4 &5752.4 &5752.4 &5752.4 &5813.6 &-61.2     \\
       &$\chi^{\sigma 3}_{1}\chi^{f2}_{m1}\chi^{c}_{m2}$  &6475.3 &       &       &       &       &        \\
       &$\chi^{\sigma 4}_{1}\chi^{f2}_{m1}\chi^{c}_{m1}$  &6166.4 &6166.4 &       &       &6194.5 &-28.1     \\
       &$\chi^{\sigma 4}_{1}\chi^{f2}_{m1}\chi^{c}_{m2}$  &6471.7 &       &       &       &       &        \\
       &$\chi^{\sigma 5}_{1}\chi^{f2}_{m1}\chi^{c}_{m1}$  &6205.0 &6205.0 &       &       &6233.2 &-28.2     \\
       &$\chi^{\sigma 5}_{1}\chi^{f2}_{m1}\chi^{c}_{m2}$  &6443.5 &       &       &       &       &        \\
       \hline
$12+$  &$\chi^{\sigma 6}_{2}\chi^{f2}_{m1}\chi^{c}_{m1}$  &6205.3 &6205.3 &6205.3 &6205.3 &6233.2 &-27.9     \\
       &$\chi^{\sigma 6}_{2}\chi^{f2}_{m1}\chi^{c}_{m2}$  &6485.6 &       &       &       &       &        \\
\hline
\end{tabular}
\end{center}
\end{table}

\begin{table}[!t]
\begin{center}
\caption{ \label{contribution2} The contributions of each terms of the Hamiltonian
for $00^+$ state in meson-meson structure in $SU(3)$ flavor symmetry (unit: MeV). }
\begin{tabular}{ccccccc} \hline
\multirow{2}*{\,} &\multicolumn{3}{c}{$0\times0 \rightarrow 0$} &\multicolumn{3}{c}{$1\times1 \rightarrow 0$}\\
\hline &$T_{bs}$  &$BK$ &$\Delta_1$    &$T_{bs}$ & $B^*K^*$ &$\Delta_2$ \\
\hline rest mass &6274.0 &6274.0 &0   &6274.0 &6274.0 &0\\
\hline
kinetic  &1607.9 &1493.4 &114.5 &1108.2 &740.3 &367.9\\
\hline
$V_{ij}^G$  &-1536.8 &-1454.1 &-82.7 &-695.1 &-488.9 &-206.2\\
\hline
$V_{ij}^C$  &-486.8 &-473.1 &-13.7 &-343.0 &-295.4 &-47.6\\
\hline
$V_{ij}^{\eta}$   &-70.2 &-65.2 &-5.0 &10.5 &3.3 &7.2\\
\hline
 $V_{ij}^{\pi}$   &-0.2 &0.0 & -0.2 &-105.6 &0.0 &-105.6\\
\hline
$V_{ij}^{K}$   &0 &0 &0 &0 &0 &0\\
\hline
$V_{ij}^{\sigma}$   &-83.2 &-0.0077 &-83.1 &-80.3 &-0.002 &-80.2\\
\hline eigenenergy  &5704.7 &5774.9 &-70.2 &6168.7 &6233.2 &-64.5\\
\hline
\end{tabular}
\end{center}
\end{table}

\begin{table}[!t]
\begin{center}
\caption{ \label{rms2} the root-mean-square(RMS) radiuses of
quarks and antiquarks of the state $00^+$ in meson-meson structure
in $SU(3)$ flavor symmetry(unit:fm). }
\begin{tabular}{ccccccc} \hline
 channel ~~~~&~~$u\bar{s}$  ~~~~&$d\bar{b}$  ~~~~&$ud$  ~~~~&$\bar{s}\bar{b}$  ~~~~&$u\bar{b}$  ~~~~&$\bar{s}d$ \\ \hline
 $0\times0 \rightarrow 0$  &0.5 &0.6 &1.9 &1.8 &1.8 &1.9\\
 \hline
 $1\times1 \rightarrow 0$  &0.7 &0.6 &1.0 &0.7 &0.8 &0.9 \\
 \hline
 coupling     &0.5 &0.6 &1.8 &1.7 &1.7 &1.8\\
 \hline
\end{tabular}
\end{center}
\end{table}

Furthermore, the root-mean-square (RMS) distances between quarks
and antiquarks in meson-meson structure for $00^+$ state are
calculated and shown in Table \ref{rms1}. For $0\times0
\rightarrow 0$ channel, the distances between the two meson
clusters are much larger than those of $u-\bar{s}$ or $d-\bar{b}$
within one cluster and it tends to be a molecular state; for
$1\times1 \rightarrow 0$ channel, the distances between the two
meson clusters are about twice of that between the quark and
antiquark in one cluster which indicates that it may be a little
compact molecular state in our present calculation. When the
coupling of two channels is considered, the dominant component of
the lowest state is $0\times0 \rightarrow 0$ color singlet-singlet
state, and the distances between the two meson clusters are a
little smaller but still far larger than that between the quark
and antiquark in one cluster, so the $00^+$ state must be a
molecular state in the present work.

The Salamanca version of the chiral quark model can describe the
meson spectrum well, where the $\sigma$ meson exchange is
considered between $u$, $d$ and $s$ quark. So it is interesting to
calculate the $T_{bs}$ in $SU(3)$ flavor symmetry, and the results
for both diquark-antidiquark structure and meson-meson structure
are demonstrated in Tables \ref{resultsu3diquark} and
\ref{resultsu3meson}, respectively. From the tables, we found that
the energies are much lower than those in the $SU(2)$ flavor
symmetry no matter in diquark-antidiquark structure or meson-meson
structure. In the diquark-antidiquark structure, the energies are
still all higher than the threshold of $BK$. In the meson-meson
structure, the energies are all below the corresponding
thresholds. For comparison, for $00^+$ state, the binding energies
are $-70.2$ MeV for spin $0\times0 \rightarrow 0$ state and
$-64.5$ MeV for spin $1\times 1 \rightarrow 0$ state, which are
much deeper than those ($-0.5$ MeV and $-10.3$ MeV) in the $SU(2)$
symmetry. From the contributions of each term of the Hamiltonian
for $00^+$ state in the $SU(3)$ symmetry (see Table
\ref{contribution2}), we can see that the $\sigma$ meson exchange
leads to the deeper binding energies for two channels.
Furthermore, the RMS distances between four particles for $00^+$
states in the $SU(3)$ flavor symmetry are demonstrated in Table
\ref{rms2}. From the table, we found that the distances between
the two meson clusters are much closer than those in the $SU(2)$
flavor symmetry due to the $\sigma$ meson exchange. For $0\times0
\rightarrow 0$ channel, it still tends to be a molecular one and
for $1\times1 \rightarrow 0$ channel, it may be a compact
tetraquark state. The effect of the channel coupling is still
tiny.

%%%%%%%%%%%%%%%%%%%%%%%%%%%%%%%%%%%%%%%%%%%%%%%%%%%%%%%%%%%%%

\section{Summary}
\label{epilogue}
Benefited from $ud\bar{s}\bar{b}$'s higher threshold than $B_s
\pi$, it has a larger mass region than $X(5568)$ to be stable and
it may be a promising detectable tetraquark state. In this paper
we try to calculate the state $ud\bar{s}\bar{b}$ ($T_{bs}$) with
quantum numbers $IJ^P(I=0,1; J=0,1,2; P=+)$ by using GEM. The
constituent chiral quark model with flavor symmetries $SU(2)$ and
$SU(3)$, which describes the light and heavy meson spectra well,
is employed in the calculation. Two structures:
diquark-antidiquark and meson-meson, are investigated. We found
that the energies of $T_{bs}$ with diquark-antidiquark structure
are all higher than the threshold of $BK$, leaving no space for
the bound state, but for the lowest energy $00^+$ state it may be
a resonance state in the $SU(2)$ flavor symmetry in our
calculation. Besides, in the $SU(2)$ flavor symmetry with the
meson-meson strucure, the mass of $00^+$ state is just below the
threshold of $BK$ with a small binding energy, $-0.6$MeV, which
can be a molecular state in the present work. As for the $SU(3)$
flavor symmetry, the results for diquark-antidiquak structure are
unaltered, and the energies with the meson-meson structure are
much lower owing to the medium-range attraction supplied by the
$\sigma$ meson exchange between $u$, $d$ and $s$ quark. From
experimental side, it is not expected that there exist so many
states, so using $\sigma$-meson exchange between $u$, $d$ and $s$
quark is not a proper way in $SU(3)$ flavor symmetry. A better way
is to employ scalar nonet in the $SU(3)$ flavor symmetry instead
of one $\sigma$ meson exchange in the $SU(2)$ flavor symmetry.

~~

\acknowledgments
This work is supported partly by the National Science Foundation
of China under Contract No. 11775118.

%%%%%%%%%%%%%%%%%%%%%%%%%%%%%%%%%%%%%%%%%%%%%%%%%%%%%%%%%%%%%%%%%%%%%%%%%%%%%%%%%%%%%%%%%%%%%%%%%%%%%%%%%%%%%%%%%%%%%%%

%%%\bibliographystyle{../../zProc/z10/z10KITPC/h-physrev4}
%%%\bibliography{../../CollectedBiB}

\begin{thebibliography}{10}
%%%%%%%%%%%%%%%%%%%%%%%%%%%%%%%%%%%%%%%%%%%%%%%%%%%%%%%%%%%%%%%%%%%
%introduction
\bibitem{Choi:2003}
S.-K. Choi \emph{et al.} (Belle Collaboration)
\newblock Phys. Rev. Lett. {\bf 91}, 262001 (2003).

\bibitem{D0Co:2016}
V.~M.~Abazov \emph{et al.} (D0 Collaboration)
\newblock Phys. Rev. Lett. {\bf 117}, 022003 (2016).

\bibitem{D0Co:2017}
V.~M.~Abazov \emph{et al.} (D0 Collaboration)
\newblock Phys. Rev. D {\bf 97}, 092004 (2018)

\bibitem{LHCb:2016}
R.~Aaij \emph{et al.} (LHCb Collaboration)
\newblock Phys. Rev. Lett. {\bf 117}, 152003 (2016).

\bibitem{CMS:2017}
A.~M.~Sirunyan \emph{et al.} (CMS Collaboration)
\newblock Phys. Rev. Lett. {\bf 120}, 202005 (2018).

\bibitem{ATLAS:2018}
M.~Aaboud \emph{et al.} (ATLAS Collaboration)
\newblock Phys. Rev. Lett. {\bf 120}, 202007 (2018).

\bibitem{CDF:2017}
T.~Aaltonen \emph{et al.} (CDF Collaboration)
\newblock Phys. Rev. Lett. {\bf 120}, 202006 (2018).

\bibitem{sumrule1} S. S. Agaev, K. Azizi and H. Sundu, Phys. Rev. D {\bf 93}, 074024 (2016).
\bibitem{sumrule2} Z. G. Wang, Commun. Theor. Phys. {\bf 66}, 335 (2016).
\bibitem{sumrule3} C. M. Zanetti, M. Nielsen and K. P. Khemchandani, Phys. Rev. D {\bf 93}, 096011 (2016).
\bibitem{sumrule4} W. Chen, H. X. Chen, X. Liu, T. G. Steele and S. L. Zhu, Phys. Rev. Let. {\bf 117}, 022002 (2016)
\bibitem{sumrule5} Liang Tang and C. F. Qiao, Eur. Phys. J. C {\bf 76}, 558 (2016).
\bibitem{qm1} W. Wang and R. L. Zhu, Chin. Phys. C {bf 40}, 093101 (2016).
\bibitem{qm2} Y. R. Liu, X. Liu and S. L. Zhu, Phys. Rev. D {\bf 93}, 074023 (2016).
\bibitem{qm3} Fl. Stancu, J. Phys. G {\bf 43}, 105001 (2016).
\bibitem{Burns} T. J. Burns, E. S. Swanson, Phys. Lett. B {\bf 760}, 627 (2016).
\bibitem{FKGuo} F. K. Guo, U. G. Meissner and B. S. Zou, Commun. Theor. Phys. {\bf 65}, 593 (2016).
\bibitem{EPJC} X. Y. Chen and J. L. Ping, Eur. Phys. J. C {\bf 76}, 351 (2016).

\bibitem{yufusheng:2017}
Fu-Sheng Yu, \newblock arXiv:1709.02571 [hep-ph].

\bibitem{Hiyama:2003cu}
E.~Hiyama, Y.~Kino and M.~Kamimura,
\newblock Prog. Part. Nucl. Phys. {\bf 51}, 223 (2003).

\bibitem{Valcarce:2005em}
A.~Valcarce, H.~Garcilazo, F.~Fernandez and P.~Gonzalez,
\newblock Rept. Prog. Phys. {\bf 68}, 965 (2005).

\bibitem{Close:2009}
Frank Close, Clark Downum,
\newblock Phys. Rev. Lett. {\bf 102}, 242003 (2009).

\bibitem{Close:2010}
Frank Close, Clark Downum, Christopher~E.~Thomas,
\newblock Phys. Rev. D. {\bf 81}, 074033 (2010).



\end{thebibliography}

\end{document}